\documentclass[12pt,a4paper]{article}
\usepackage{epsfig,amssymb,amsbsy}

\font\got=eufm10 at 12pt
\font\sgot=eufm10 at 10pt

\newtheorem{theorem}{Theorem}[section] 
\newtheorem{lemma}[theorem]{Lemma}
\newtheorem{proposition}[theorem]{Proposition}

\newtheorem{remark}[theorem]{Remark}
\newtheorem{example}[theorem]{Example}

\def\balpha{\boldsymbol\alpha}
\def\bbeta{\boldsymbol\beta}
\def\build#1_#2^#3{\mathrel{\mathop{\kern 0pt#1}\limits_{#2}^{#3}}}
\def\c{{\mathcal C}}
\def\C{\mathbb{C}}
\def\H{\mathbb{H}}
\def\G{\Gamma}
\def\epsilon{\varepsilon}
\def\L{\rm L}
\def\End{{\mathop{\rm End}}}
\def\Hom{{\mathop{\rm Hom}}}
\def\id{\mathop{\rm Id}}
\def\lra{\longrightarrow}
\def\o{{\cal O}}
\def\p{{\cal P}}
\def\R{\mathbb{R}}
\def\S{\hbox{\got{S}}}
\def\sS{\hbox{\sgot{S}}}
\def\Sym{{\rm Sym}}
\def\tr{\mathop{\rm tr}}
\def\pf{\textsc{Proof}}
\def\qed{\hfill\hbox{\vrule\vbox to 2mm{\hrule width 2mm\vfill\hrule}\vrule}
  \\} 

\title{Wilson loops in the light of spin networks}

\author{Thierry L\'evy \thanks{IRMA -- 7, rue Ren\'e Descartes -- F-67084
Strasbourg Cedex}}

\begin{document}
\maketitle

\begin{abstract}
If $G$ is any finite product of orthogonal, unitary and symplectic
matrix groups, then Wilson loops generate a dense subalgebra of
continuous observables on the configuration space of lattice gauge
theory with structure group $G$. If $G$ is orthogonal, unitary or
symplectic, then Wilson loops associated to the natural representation
of $G$ are enough. 

This extends a result of A. Sengupta \cite{Sen}. In particular, our
approach includes the case of even orthogonal groups.
\end{abstract}

\section{Introduction}

On a compact Lie group, the Peter-Weyl theorem asserts that the
characters of irreducible representations generate a dense subalgebra
of continuous functions invariant by adjunction. In lattice gauge
theory, configuration spaces are powers of a Lie group on which
another power of the same group acts, according to the geometry of a
given graph and in a way which extends the adjoint action of the group
on itself. Peter-Weyl theorem can be adapted to this situation and
the functions that play the role of the characters are
called {\it spin networks}. Despite the fact that spin
networks were introduced about forty years ago in a physical
context\footnote{R. Penrose introduced them for the purposes of
quantization of the geometry of space. See \cite{Smolin} for a
historical account.}, their importance in lattice gauge theory has been
recognized rather recently \cite{Baez}. In the mean time, another set
of functions, easier to define, has been used as the standard set of
observables: Wilson loops. However, it is not clear at all a priori
that this set is complete, that is, that Wilson loops generate a dense
subalgebra of continuous invariant functions on the configuration
space. A. Sengupta has proved in \cite{Sen} that it is true when the
group is a product of odd orthogonal, unitary (and
symplectic) groups. In this paper, an approach similar to that
of Sengupta but with a little more classical invariant theory combined
with the use of spin networks allows us to add even orthogonal groups
to the list and, hopefully, to clarify the argument. 

The problem of completeness of Wilson loops can be expressed in three
equivalent ways. The first one is described above. The second one is
more geometrical and consists in asking whether a connection on a
principal bundle is determined up to gauge transformation by the
conjugacy classes of its loop holonomies. The third one is more
algebraic: is it true that the diagonal conjugacy class of a finite collection
of elements of a compact Lie group is determined by the conjugacy
classes of all possible products one may form with these elements and their
inverses ? The equivalence of these questions is discussed in
\cite{Sen}, and we will make an important use of the equivalence
between the first and the third point of view.   


\section{The configuration space}

Let $G$ be a compact connected Lie group. Let $\G=(E,V)$ be a graph
with oriented edges. By this we mean that $V$ is a finite set and $E$
is a set of pairs of elements of $V$. Diagonal pairs are allowed and a
pair can occur several times in $E$. If $e=(v,w) \in E$ is an edge, we
define the source and target of $e$ respectively by $s(e)=v$ and
$t(e)=w$. We make the assumption that no vertex is isolated, that is,
$s(E)\cup t(E)=V$. 

Define an action of $G^V$ on $G^E$, as follows. For
$\phi=(\phi_v)_{v\in V} \in G^V$ and $g=(g_e)_{e\in E} \in G^E$, set
$$\phi \cdot g=\left((\phi \cdot g)_e\right)_{e\in E} \;\;\; {\rm with}
\;\;\; (\phi \cdot g)_e=\phi_{t(e)}^{-1} g_e \phi_{s(e)}.$$

The configuration space for lattice gauge theory on $\Gamma$ with
structure group $G$ is the topological quotient space
$\c_\G^G=G^V\backslash G^E$ and it can be thought of as a
finite-dimensional approximation of a space of connections modulo
gauge transformations.  

\begin{example}\rm 
Consider the very simple graph $\L_1$ with one single vertex $v$ and
one single edge $(v,v)$. Then $\c^G_{\L_1}$ is just the space of conjugacy
classes on $G$. 
\end{example}

\begin{example}\rm Choose an integer $r\geq 1$ and consider the graph
$\L_r$ with $r$ edges depicted below.

\begin{figure}[h]
\begin{center}
\includegraphics{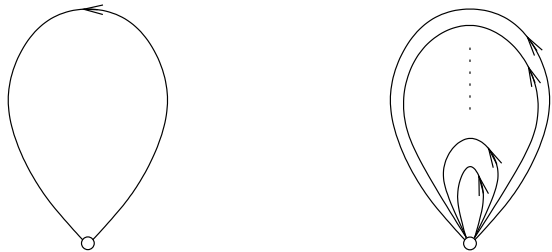}
\caption{The graphs $L_1$ and  $\L_r$.}
\end{center}
\end{figure}

For this graph, $G^E=G^r$ on which $G^V=G$ acts by diagonal
conjugation, and we will call {\it diagonal conjugacy classes} of $G^r$ the
points of $\c^G_{\L_r}$. 
\end{example}

\begin{remark}\rm
If $\Gamma$ is a tree, you may check that $\c^G_\Gamma$ is a single
point.
\end{remark}

Wilson loops are continuous functions on $\c^G_\Gamma$ or,
equivalently, continuous functions on $G^E$ invariant under the action
of $G^V$. We recall briefly how they are defined. 

Let $E^\pm$ denote the set containing twice each edge of $\Gamma$,
once with its natural orientation and once with the reversed
one. Formally, set $E^\pm=E\times \{+,-\}$, extend the functions $s$
and $t$ to $E^\pm$ by $s(e,+)=s(e)$, $s(e,-)=t(e)$ and the two similar rules
for $t$. A point of $G^E$ determines a point of $G^{E^\pm}$ by the
rules $g_{(e,+)}=g_e$ and $g_{(e,-)}=g_e^{-1}$. For the sake of
clarity, we identify $e$ with $(e,+)$ and denote $(e,-)$ by
$e^{-1}$. Moreover, we use the notation $e$ to denote a generic
element of $E^\pm$.

A path in $\Gamma$ is a finite sequence $p=(e_1,\ldots,e_n)$ of
elements of $E^\pm$ such that $t(e_i)=s(e_{i+1})$ for all
$i=1,\ldots,n-1$. It is a loop based at $v$ if $t(e_n)=s(e_1)=v$. To
a loop $l=(e_1,\ldots,e_n)$ one associates a 
function $h_l: G^E \lra G$ defined by $h_l(g)=g_{e_n}\ldots
g_{e_1}$. One checks easily that the action of $\phi \in G^V$ on $G^E$
conjugates $h_l$ by $\phi_{s(e_1)}^{-1}$ so that, given any
finite-dimensional representation $\alpha$ of $G$ with character
$\chi_\alpha$, the function  
$$W_{\alpha,l}=\chi_\alpha \circ h_l : \c^G_\Gamma \lra \C$$
is well-defined. It is called a Wilson loop. 

\begin{remark} \rm  A wider class of functions can be defined
on $\c^G_\G$. Instead of considering one loop, we can consider several
loops $l_1,\ldots,l_n$ based at the same point. Then, for any function
$f:G^n \lra \C$ invariant by diagonal adjunction, that is, such that
for all $g_1,\ldots,g_n,h \in G$, one has
$f(g_1,\ldots,g_n)=f(hg_1h^{-1},\ldots,hg_nh^{-1})$, the function 
$$f\circ (h_{l_1},\ldots,h_{l_n}) : \c^G_\G \lra \C$$
is well-defined. In words, the diagonal conjugacy
class of $(h_{l_1}(c),\ldots,h_{l_n}(c))$ is well-defined for every $c$
in the configuration space. 
\end{remark}

\section{Statement of the result}

In this paper, $O(n)$ and $SO(n)$ denote respectively the groups $O_n \R$ and
$SO_n \R$. By the symplectic group $Sp(n)$ we mean the
subgroup\footnote{$Sp_{2n}\C$ is the group of
matrices which preserve the skew-symmetric form whose matrix in the
canonical basis is $\scriptsize{ \pmatrix{0 & I \cr -I & 0}}$.} $U(2n)
\cap Sp_{2n}\C$ of $GL_{2n}\C$. It is isomorphic to
the quaternionic unitary group $U_\H(n)$. The main result is the following. 

\begin{theorem} \label{main} Let $G$ be a finite product of groups
among $U(n)$, $SU(n)$, $O(n)$, $SO(n)$, $Sp(n)$. Let
$\Gamma=(E,V)$ be a 
graph. Then the algebra generated by the Wilson loops is dense in the
space of continuous functions on $\c^G_\G=G^E/G^V$.
\end{theorem}

\begin{example}\rm In the case of the graph $\L_1$, Theorem \ref{main}
is equivalent to Peter-Weyl theorem.
\end{example}

\begin{example}\rm Consider the case of the graph $L_r$. Loops in
$\L_r$ are in one-to-one correspondence with words in the letters of
$E^\pm=\{e_1^{\pm 1},\ldots,e_r^{\pm 1}\}$. For such a word $w$ and
given a point $g=(g_1,\ldots,g_r)$ of $G^E$, let us denote by $w(g)$ the
corresponding product in reversed order of the $g_i's$ and their
inverses. Observe that, if a loop $l$ corresponds to a word $w$, then
$h_l(g)=w(g)$ for all $g$. 

Assume for a moment that Theorem \ref{main} is proved for the graphs
$\L_r$. We can rephrase it as follows.
\end{example}

\begin{proposition} \label{Ln-alg} Let $G$ be a group as in Theorem
\ref{main}. If $g$ and $g'$ are two points of
$G^r$ such that 
for all word $w$ in $r$ letters and their inverses, the elements
$w(g)$ and $w(g')$ of $G$ are conjugate, then $g$ and $g'$ belong to
the same diagonal conjugacy class.
\end{proposition}

\pf. In this proof, we identify freely $G^r$ with $G^E$, where $E$ is
the set of edges of the graph $\L_r$. If two points $g$ and $g'$ of 
$G^r$ do not belong to the same diagonal conjugacy class, their orbits
in $\c^G_{\L_r}$ are different. Hence, by Theorem \ref{main} applied to
the graph $\L_r$, there exists a loop $l$ in $\L_r$ such that $h_l(g)$
and $h_l(g')$ are not conjugate. This loop is a word $w$ in the
letters of $E^\pm$ and the corresponding elements $w(g)$ and $w(g')$,
which are precisely $h_l(g)$ and $h_l(g')$, are not conjugate.\qed

It turns out that Proposition \ref{Ln-alg} is almost equivalent to
Theorem \ref{main}. The gap is filled by the following result.

\begin{proposition} \label{fill-gap} Let $G$ be a compact group. Let
$\G=(E,V)$ be a 
graph. Let $c$ and $c'$ be two points of $\c^G_\G$. Assume that,
for any vertex $v$ of $V$ and any finite sequence $l_1,\ldots,l_r$ of
loops in $\G$ based at $v$, the diagonal conjugacy classes of
$(h_{l_1}(c),\ldots,h_{l_r}(c))$ and
$(h_{l_1}(c'),\ldots,h_{l_r}(c'))$ are equal. Then $c=c'$. 
\end{proposition}

This proposition is proved in a slightly different language
in \cite{Sen_S2}. For the convenience of the reader, we recall the
argument.\\

\pf. Fix once for all a vertex $v$. Choose $g$ and $g'$ in $G^E$
representing $c$ and $c'$. For any finite family $F$ of loops
based at $v$, let $K_F$ be the closed subset of $G$ consisting of those $k$
such that $h_l(g')=k h_l(g) k^{-1}$ for all $l \in F$. By assumption,
$K_F$ is non-empty, just as any finite intersection of sets of the
form $K_F$. By compactness of $G$, there exists $k$ such that
$h_l(g')=k h_l(g) k^{-1}$ for every loop $l$ based at $v$. By letting
the element of $G^V$ equal to $k$ at $v$ and 1 anywhere else act on
$g'$, we are reduced to the case where $h_l(g)=h_l(g')$ for all $l$
based at $v$.

Now, for every vertex $w$, choose a path $p$ in $\G$ joining $w$ to
$v$. Define $\phi_w=h_p(g)h_{p^{-1}}(g')$. Then one checks easily that
$\phi_w$ does not depend on $p$ and that the element
$\phi=(\phi_w)_{w\in V}$ of $G^V$ built in that way satisfies $\phi\cdot g
=g'$. Hence, $c=c'$. \qed

We have reduced the problem as follows.

\begin{proposition} Theorem \ref{main} is logically equivalent to its
specialization to the graphs $\L_r, r\geq 1$, which is in turn
equivalent to Proposition \ref{Ln-alg}.
\end{proposition}

\pf. We prove that Proposition \ref{Ln-alg} implies Theorem \ref{main}. Let
$\Gamma$ be a graph. Let $g$ and $g'$ be two points of $G^E$ 
such that all Wilson loops take the same value at $g$ and $g'$. Let
$v$ be a vertex of the graph and $l_1,\ldots,l_r$ $r$ loops based at
$v$. Since any product of the $l_i$'s and their inverses is still a
loop based at $v$, Proposition \ref{Ln-alg} applied to the elements
$(h_{l_1}(g),\ldots,h_{l_r}(g))$ and
$(h_{l_1}(g'),\ldots,h_{l_r}(g'))$ of $G^r$ shows that there exists $k\in G$
such that $h_{l_i}(g')=kh_{l_i}(g)k^{-1}$ for all $i=1\ldots r$. 
Hence, by Proposition \ref{fill-gap}, $g$ and $g'$ belong to the same
orbit under the action of $G^V$. Hence, Wilson loops separate the
points on the configuration space. Since this space is compact, the
result follows by the Stone-Weierstrass theorem. \qed

The translation in algebraic language allows us to reduce the list of
groups that we need to consider. The proof of the following lemma is
straightforward.

\begin{lemma} If Proposition \ref{Ln-alg} holds for two groups $G_1$
and $G_2$, then it holds for their product $G_1\times G_2$.
\end{lemma}

According to this lemma, it is enough to prove Theorem
\ref{main} when $G$ is one of the groups $O(n)$, $SO(n)$, $U(n)$,
$SU(n)$, $Sp(n)$. \\

\begin{remark} {\rm One might expect that the property expressed by
Proposition 
\ref{Ln-alg} is preserved by standard transformations of the group
such as quotients or central extensions. Unfortunately, no such result
seems easy to prove. For central extensions, A. Sengupta 
has stated and proved in \cite{Sen} a partial result, namely that a property
slightly stronger than that of Proposition \ref{Ln-alg} is
preserved. I have not been able to improve this result.}
\end{remark}

\section{Spin networks}

From now on, we concentrate on the case where $\Gamma$ is the graph
$L_r$ for some $r\geq 1$ and $G$ is one of the groups listed
above. Instead of working on the configuration space, we prefer to
work on $G^E=G^r$ and consider only objects which are invariant under
the diagonal adjoint action of $G$.  

Spin networks provide us with a very natural dense subalgebra
of the space of invariant continuous functions. They are defined
as follows.  

Choose $r$ finite-dimensional representations $\alpha_1,\ldots,\alpha_r$ of
$G$ with spaces $V_1,\ldots,V_r$. Then $G$ acts on $V_1\otimes \ldots
\otimes V_r$ by $\alpha_1\otimes \ldots \otimes \alpha_r$. Let us
choose $I \in \End_G(V_1\otimes \ldots \otimes V_r)$. This means that
$I$ is a linear endomorphism of $V_1\otimes \ldots \otimes V_r$ commuting with
the action of $G$. Let $g$ be an element of $G^r$. Set
${\balpha}=(\alpha_1,\ldots,\alpha_r)$. Then the function  
$\psi_{\balpha,I} : G^r \lra \C$
defined by
$$\psi_{\balpha,I}(g)= \tr (\alpha_1(g_1) \otimes \ldots \otimes
\alpha_r(g_r) \circ I)$$
is invariant under the action of $G$. It is called a {\it spin
network}.

The following proposition has been proved by J. Baez \cite{Baez}. 

\begin{theorem} \label{spin-net} The spin networks $\psi_{\balpha,I}$,
where $\balpha$ 
runs over the set of $r$-tuples of irreducible representations of $G$
and, given $\balpha=(\alpha_1,\ldots,\alpha_r)$, $I$ runs over a basis
of $\End_G(V_1\otimes \ldots \otimes V_r)$, generate a dense
subalgebra of $C(G^r)^G$, the space of continuous functions invariant
under the diagonal action of $G$.
\end{theorem}

\begin{remark} \rm Just as in Peter-Weyl theorem, there is also a
$L^2$ version of this result, but we do not need it here. 
\end{remark}

For the sake of completeness and because we find it illuminating, we
give a short proof of Theorem \ref{spin-net}.\\

\pf. The irreducible representations of $G^r$ are exactly the tensor
products of $r$ irreducible representations of $G$. Thus, Peter-Weyl
theorem applied to $G^r$ implies that the functions
$\psi_{\balpha,J}$ on $G^r$, where $\balpha$ is as before, but $J$ is
{\it any} endomorphism of $V_1\otimes \ldots \otimes V_r$, generate a
dense subalgebra of $C(G^r)$. 

Now, it is readily seen that the average under the diagonal action of
$G$ of such a function $\psi_{\balpha,J}$ is a spin network
$\psi_{\balpha,I}$, where $I$ is the orthogonal projection of $J$ on
$\End_G(V_1\otimes \ldots \otimes V_r)$ for any $G$-invariant scalar
product on $\End(V_1\otimes \ldots \otimes V_r)$. The result follows
immediately. \qed


Let us call the spin network $\psi_{\balpha,I}$ {\it
irreducible}  if $\balpha$ is irreducible as a representation of
$G^n$, that is, if every $\alpha_i$ is irreducible. 

\begin{proposition} \label{spin-irred} Any spin network is a linear
combination of irreducible spin networks.
\end{proposition}

\pf. Let $\psi_{\balpha,I}$ be a spin network. Decompose $\balpha$ as
a sum $\bigoplus_k \balpha_k$ of irreducible representations of
$G^n$. Accordingly, decompose the space $V$ of $\balpha$ as
$V=\bigoplus_k V_k$. For each $k$, define $I_k$ as the component of
$I$ lying in $\End_G(V_k)$ in the decomposition
$$\End_G(V)=\End_G(\bigoplus_k V_k) \simeq \bigoplus_{k,l}
\Hom_G(V_k,V_l).$$
Then we leave it for the reader to check that $\displaystyle
\psi_{\balpha,I}=\sum_k \psi_{\balpha_k, I_k}. $ \qed


In order to establish Theorem \ref{main} for the graphs $\L_r$, it is
thus enough to prove the following result.

\begin{proposition}\label{spin-wilson-ln} Let $G$ be one of the groups
$O(n)$, $SO(n)$, $U(n)$, $SU(n)$, $Sp(n)$. Let $r\geq 1$ be an
integer. On the graph $\L_r$, any irreducible spin network is a finite
linear combination of products of Wilson loops. 
\end{proposition}

We have now almost reached the formulation of the problem under which we are
going to solve it. 

\section{Natural representations}

The main problem we are going to encounter in handling with spin
networks is that they involve invariant endomorphisms of spaces of
representations of $G$, which are in general very difficult to
describe. 

In the case where $G$ is a group of complex matrices of some size $n$, that
is, an orthogonal, unitary or symplectic group\footnote{Recall that
the elements of $Sp(n)$ are complex matrices of size $2n$.}, $G$ acts by left
multiplication on $V=\C^n$ and this is called the {\it natural
representation}. The contragredient of this representation is the
action on $V^*$ given by $g\cdot \varphi=\varphi \circ g^{-1}$. 

The {\it first fundamental theorems} (FFT) of classical invariant
theory describe a set of generators of the space $\End_G(V^{\otimes p} \otimes
(V^*)^{\otimes q})$ when $p$ and $q$ are given integers, for the
different kinds of matrix groups $G$.  
  
This gives us what we are looking for in a special case, namely when
each representation $\alpha_i$ is of the form $V^{\otimes p} \otimes
(V^*)^{\otimes q}$. The two following results allow us to
reduce the general case to this particular one.

\begin{lemma}\label{submodule} Let $G$ be any compact Lie group. Consider
$\balpha=(\alpha_1,\ldots,\alpha_r)$ and
$\bbeta=(\beta_1,\ldots,\beta_r)$ two $r$-tuples of representations of
$G$. Assume that, for each $i=1\ldots r$, the representation
$\alpha_i$ is a subrepresentation of $\beta_i$. Let $\psi_{\balpha,I}$
be a spin network on $G^r$. Then there exists $J \in
\End_G(\beta_1\otimes\ldots\otimes \beta_r)$ such that
$\psi_{\balpha,I}$=$\psi_{\bbeta,J}$.
\end{lemma}

\pf. For each $i$, endow the space of $\beta_i$ with a $G$-invariant
scalar product and define $p_i$ as the orthogonal projection on a
subspace on which the action of $G$ is isomorphic to $\alpha_i$. Then
$J=I \circ p_1\otimes \ldots \otimes p_r$
is $G$-invariant and satisfies $\psi_{\balpha,I}$=$\psi_{\bbeta,J}$. \qed

\begin{proposition} \label{anyrep-nat} Let $G$ be a compact Lie
group. Let $\alpha$ be a faithful finite-dimensional representation of
$G$. Then any irreducible representation of $G$ is a
subrepresentation of $\alpha^{\otimes p}\otimes (\alpha^\vee)^{\otimes
q}$ for some integers $p,q \geq 0$.
\end{proposition}

In this statement, $\alpha^\vee$ denotes the contragredient
representation of $\alpha$. We use the convention $\alpha^{\otimes
0}=\C$, the trivial representation. 

This result is of course well-known\footnote{In the case of a finite
group, it is referred to as a theorem of Burnside and Molien in
\cite{FH}.} in the sense that the
representations of compact Lie groups are completely classified and
that a proof ``by inspection'' is almost possible, see for example the
end of \cite{BtD}. However, we were not able to find a direct proof
in textbooks on Lie groups. Therefore, we propose a short analytical
argument.\\  

\pf. Let $\alpha$ be a faithful finite-dimensional representation of
$G$. Since $\alpha$ 
is unitary for some Hermitian scalar product, its character satisfies the
inequality $|\chi_\alpha(g)| \leq \chi_\alpha(1)$ with equality only
if $\alpha(g)=\pm \id$. Hence, $|\chi_\alpha(g)+1|$ is maximal only
when $\alpha(g)=\id$, that is, since $\alpha$ is faithful, when
$g=e$, the identity element of $G$. This implies immediately that the
probability measures
$$\mu_n=\frac{|\chi_\alpha(g)+1|^{2n}}{\int_G |\chi_\alpha(g)+1|^{2n} \; dg}
\; dg$$
on $G$ converge weakly to the Dirac mass $\delta_e$. Here, $dg$
denotes the unit-mass Haar measure on $G$. In particular, let
$\rho$ be any irreducible representation of $G$. Since
$\mu_n(\chi_\rho)$ converges to $\chi_\rho(e) \neq 0$, there exists an
integer $n\geq 0$ such that
$$\int_G \chi_\rho(g)  |\chi_\alpha(g)+1|^{2n} \; dg \neq 0.$$
Now observe that $|\chi_\alpha+1|^2$ is just the character of the
representation $(\alpha\oplus \C)\otimes(\alpha\oplus \C)^\vee \simeq
\C \oplus \alpha \oplus \alpha^\vee \oplus (\alpha \otimes
\alpha^\vee)$, where $\C$ denotes the trivial representation of
$G$. Thus, $\rho$ is a subrepresentation of the $n$-th tensor 
product of this representation. This tensor product breaks into
(non-necessarily irreducible) factors of the form $\alpha^{\otimes 
p}\otimes (\alpha^\vee)^{\otimes q}$, so that $\rho$, being
irreducible, is a subrepresentation of one of them. \qed

\begin{remark}\rm We have not used the fact that $G$ was a Lie group,
we have only used its compactness. However, a compact group admits a
faithful finite-dimensional representation if and only if it is a Lie
group (see for example \cite{Rob}). 
\end{remark}

For matrix groups, Proposition \ref{anyrep-nat} ensures that every
irreducible representation arises as a subrepresentation of some
tensor product of a number of copies of the natural representation and
its contragredient. We are now reduced to prove the following result.

\begin{proposition}\label{spin-wilson-ln-nat} Let $G$ be a group of
the following list: $O(n)$, 
$SO(n)$, $U(n)$, $SU(n)$, $Sp(n)$. Let $r\geq 1$ be an integer. Let
$\balpha$ be a $r$-tuple of representations of the form $V^{\otimes
p} \otimes (V^*)^{\otimes q}$, where $V$ is the natural representation
of $G$. Then any spin network $\psi_{\balpha,I}$ on $G^r$ is a linear
combination of products of Wilson loops.
\end{proposition}

We leave it to the reader to check that Proposition
\ref{spin-wilson-ln-nat} implies Propostion \ref{spin-wilson-ln}. \\


\section{Unitary groups}

Let $n\geq 1$ be an integer and let $G$ be either $U(n)$ or
$SU(n)$. The group $G$ acts on $V=\C^n$ by multiplication on the left. For any
integer $d\geq 1$, there is a corresponding diagonal action of $G$ on 
$V^{\otimes d}$, that we denote by $\rho : G\lra GL(V^{\otimes
d})$. On the other hand, the symmetric group $\S_d$ acts by 
permutation of the factors on $V^{\otimes d}$. We denote this action
by $\pi : \S_d \lra GL(V^{\otimes d})$. It is obvious that 
the actions $\rho$ and $\pi$ commute to each other. The following theorem is
known as Schur-Weyl duality theorem.

\begin{theorem}[Schur-Weyl duality] The two subalgebras $\rho(\C U(n))$
and $\pi(\C \S_d)$ of $\End(V^{\otimes d})$ are each other's
commutant.
\end{theorem}

In other words, $\End_{U(n)}(V^{\otimes d})$ is generated as a vector space
by the permutations of the factors. The case of $SU(n)$ follows
immediately, since $\End_{SU(n)}(V^{\otimes d})=\End_{U(n)}(V^{\otimes d})$.\\

\pf. By the bicommutant theorem (see \cite{Lang} for example), it is
equivalent to prove that 
$\pi(\C \S_d)'= \rho(\C G)$ or to prove that  $\rho(\C G)'=\pi(\C
\S_d)$. The second statement is the most important for us, but the
first one is the easiest to prove. 

By definition, $\pi(\C \S_d)'=\End_{\sS_d}(V^{\otimes d})$, which in
turn is just $\End(V^{\otimes d})^{\sS_d}$, where $\S_d$ acts by
conjugation on $\End(V^{\otimes d})$. Now,
$$\End(V^{\otimes d})^{\sS_d} \simeq \left[\End(V)^{\otimes
d}\right]^{\sS_d} \simeq \Sym^d (\End(V)).$$
We must prove that $\Sym^d( \End(V))$ is
generated by the endomorphisms of the form $\rho(g)^{\otimes d}$,
$g\in U(n)$. This is true because $U(n)$ is Zariski-dense in
$\End(\C^n)$ and, for any finite-dimensional vector space $W$, $\Sym^d(W)$ is
generated by $\{ x^{\otimes d} \; | \; x\in X\}$ as soon as $X$ is
Zariski-dense\footnote{This is most easily seen through the
identification $\Sym^d (W) \simeq \p^d(W)^*$, where
$\p^d$ denotes the algebra of homogeneous polynomials of degree $d$.}
in $W$. \qed

Consider the following isomorphisms of $G$-modules:
\begin{eqnarray}
\End(V^{\otimes p} \otimes (V^*)^{\otimes q}) &\simeq&
(V^*)^{\otimes p} \otimes V^{\otimes p} \otimes V^{\otimes q} \otimes
(V^*)^{\otimes q} \nonumber \\
& \simeq &  (V^*)^{\otimes p+q} \otimes V^{\otimes
p+q} \nonumber \\
&\simeq  & \End(V^{\otimes p+q}), \label{SW-isom}
\end{eqnarray}
where the second one is chosen in the simplest possible way, namely
$$\varphi_1  \ldots  \varphi_p   u_1  
\ldots  u_p   v_1   \ldots  v_q  
\psi_1  \ldots  \psi_q \mapsto  \varphi_1 
\ldots  \varphi_p   \psi_1  \ldots  \psi_q   u_1  
\ldots  u_p   v_1   \ldots  v_q .$$

If $\sigma$ belongs to $\S_{p+q}$, let us denote by $I_\sigma$ the
element of $\End(V^{\otimes p} \otimes (V^*)^{\otimes q})$
corresponding via (\ref{SW-isom}) to $\pi(\sigma)$. Schur-Weyl duality
implies that $\End_G(V^{\otimes p} \otimes (V^*)^{\otimes q})$ is
generated by the endomorphisms $I_\sigma$. 

Let $p_1,\ldots,p_r,q_1,\ldots,q_r$ be
non-negative integers. For each $i=1\ldots r$, consider the representation
$\alpha_i=V^{\otimes p_i} \otimes (V^*)^{\otimes q_i}$ of $G$ and set
$\balpha=(\alpha_1,\ldots,\alpha_r)$. Set $p=p_1+\ldots+p_r$ and
$q=q_1+\ldots+q_r$. Let $\sigma$ be an element of $\S_{p+q}$. Consider
$I_\sigma \in \End_G(V^{\otimes p} \otimes (V^*)^{\otimes q})$. By the
isomorphism $V^{\otimes p} \otimes (V^*)^{\otimes q}\simeq
\otimes_{i=1}^r (V^{\otimes p_i} \otimes (V^*)^{\otimes q_i})$,
$I_\sigma$ can be seen as an element of $\End_G(\alpha_1\otimes
\ldots\otimes \alpha_r)$. We may thus form the spin network
$\psi_{\balpha,I_\sigma}$. The following proposition implies
Proposition \ref{spin-wilson-ln-nat} in the case of unitary groups.

\begin{proposition} \label{core-unitary} The spin network
$\psi_{\balpha,I_\sigma}$ on $G^r$ is a product of Wilson loops.
\end{proposition}

\pf. Let us denote by $n$ the natural representation of $G$ and
$n^\vee$ its contragredient. By definition, 
$$\psi_{\balpha,I_\sigma}(g_1,\ldots,g_r)=\tr\left(\bigotimes_{i=1}^r
n(g_i)^{\otimes p_i} \otimes n^\vee(g_i)^{\otimes q_i} \circ
I_\sigma\right).$$
Each factor $n(g_i)^{\otimes p_i} \otimes n^\vee(g_i)^{\otimes q_i}$
corresponds, through (\ref{SW-isom}) with $p=p_1$ and $q=q_1$, to
$n(g_i)^{\otimes p_i} \otimes n(g_i^{-1})^{\otimes q_i}$, by
definition of the contragredient. Thus, 
$$\psi_{\balpha,I_\sigma}(g_1,\ldots,g_r)=\tr\left(\bigotimes_{i=1}^r
n(g_i)^{\otimes p_i} \otimes \bigotimes_{i=1}^r n(g_i^{-1})^{\otimes
q_i} \circ \pi(\sigma) \right),$$
where we see now both endomorphisms as elements of $\End(V^{\otimes
p+q})$. This trace can now easily be evaluated. Before that and for
the sake of 
clarity, let us rename the sequence
$(g_1,\ldots,g_1,\ldots,g_r,\ldots,g_r,
g_1^{-1},\ldots,g_1^{-1},\ldots,g_r^{-1},\ldots,g_r^{-1})$,   
where $g_i$ appears $p_i$ 
times and $g_i^{-1}$ $q_i$ times, as $(h_1,\ldots,h_{p+q})$. Then
the tensor product appearing in the last equation is just $h_1\otimes
\ldots \otimes h_{p+q}$. Hence, 
$$\psi_{\balpha,I_\sigma}(g_1,\ldots,g_r)=\prod_{\c=(a_1\ldots a_k)}
\tr(h_{a_1}\ldots h_{a_k}),$$
where the product runs over the cycles of $\sigma$. We claim that each
factor in this product is a Wilson loop. To see this, define the
functions $j:\{1,\ldots,p+q\} \lra \{1,r\}$ by
\begin{eqnarray*} j(a)=i \;\; &{\rm if }& \;\; p_1+\ldots+p_{i-1}<a\leq
p_1+\ldots+p_i \\
&{\rm or }& \;\; p+q_1+\ldots+q_{i-1}<a\leq p+q_1+\ldots+q_i
\end{eqnarray*}
and $\epsilon: \{1,\ldots,p+q\} \lra \{1,-1\}$ such that $\epsilon(a)$ is
$+1$ if $1\leq a\leq p$ and $-1$ if $p+1\leq a\leq q$. They are
designed in such a way that $h_a=g_{j(a)}^{\epsilon(a)}$.

Let us now give a name to the edges of the graph $\L_r$, namely set
$E=(e_1,\ldots,e_r)$. For each cycle $\c=(a_1\ldots a_k)$ of $\sigma$,
define a loop $l_\c$ in $\L_r$ by $l_\c=(e_{j(a_k)}^{\epsilon(a_k)},
\ldots, e_{j(a_1)}^{\epsilon(a_1)})$. Then the last equality can be
rewritten simply as
$$\psi_{\balpha,I_\sigma}(g_1,\ldots,g_r)=\prod_{\c=(a_1\ldots a_k)}
W_{n,l_\c}(g_1,\ldots,g_r)$$
and the result is proved. \qed

This proof has a nice graphical representation which allows one to
understand very easily the generalization to the orthogonal and
symplectic cases. 

\begin{figure}[h]
\begin{center}
\scalebox{0.7}{\includegraphics{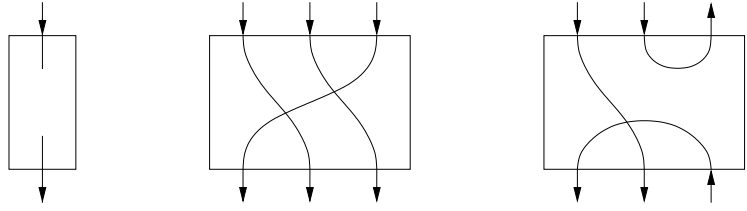}}
\caption{\label{tens}Schematic representation of tensors.}
\end{center}
\end{figure}

Let us represent a tensor of $V^{\otimes p}\otimes (V^*)^{\otimes q}$
by a box with $p+q$ oriented legs, $p$ outwards and $q$ inwards. We put inside
the box a schematic description of the tensor. For example, the
leftmost picture in figure \ref{tens} represents a tensor of
$V^*\otimes V$. It could be labeled by an element of $\End(V)$ or
$\End(V^*)$.

The middle picture represents the tensor $\pi((123)) \in
\End(V^{\otimes 3})$. The rightmost picture represents the same
tensor, via the identification\footnote{We will stay a bit loose about
the order of the factors in the tensors. We hope the pictures are
clear enough by themselves.} $\End(V^{\otimes 3})\simeq \End(V^{\otimes
2}\otimes V^*)$.

In this representation, tensor product corresponds to juxtaposition
of the boxes and a contraction is represented by joining an outcoming
leg with an incoming one.

\begin{figure}[h]
\begin{center}
\include{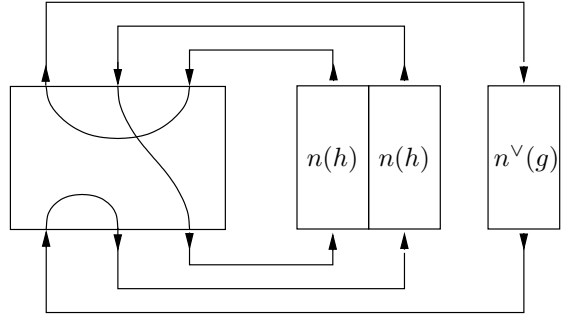}
\caption{\label{loop}The spin network $\psi_{(n^\vee,n^{\otimes
2}),I_{(123)}}$ on 
$\L_2$ as the Wilson loop $W_{n,(e_1^{-1},e_2,e_2)}$.}
\end{center}
\end{figure}

Let us consider a particular case, for example $r=2$, $p_1=q_2=0$, $q_1=1$ and
$p_2=2$. We take the permutation $\sigma=(123)$. Choose $(g,h)
\in G^2$. The picture corresponding to $\tr(n^\vee(g) \otimes
n(h)^{\otimes 2} \circ I_\sigma)$ is drawn below (Figure \ref{loop}).

If one remembers that, through the identification $\End(V^*)\simeq
\End(V)$, $n^\vee(g)$ corresponds to $n(g^{-1})$, it becomes almost
evident that the trace we are computing is also a Wilson loop, namely
$\tr n(g^{-1}h^2)$. 

\section{Orthogonal and symplectic groups}

Let $n\geq 1$ be an integer. Let $G$ be either $O(n)$, $SO(n)$ or
$Sp(n)$. Recall that, by the two first groups of this list we mean
respectively $O_n \R$ and $SO_n \R$. By the third we mean the subgroup
$U(2n) \cap 
Sp_{2n}\C$ of $GL_{2n}\C$, which preserves, via the identification
$\H^n\simeq \C^n\oplus j\C^n$, the standard quaternionic Hermitian
scalar product on $\H^n$. We are going to treat at once the orthogonal
and symplectic case, although they are not exactly identical. For
example, the space $V$ of the natural representation of $G$ is $\C^n$
in the orthogonal case, $\C^{2n}$ in the symplectic case. We shall use
the letter $m$ to denote the dimension of $V$ in both cases. In the
orthogonal case, we are going to use orthonormal bases of $V$. In the
symplectic case, we say that $(e_1,\ldots,e_{2n})$ is a standard basis
for $V$ if $\langle e_i,e_{i+n} \rangle =1$ for $i=1,\ldots,n$ and
$\langle e_i,e_j \rangle=0$ if $|i-j|\neq n$. 

The situation here differs from the preceding one in two main respects
because $G$ preserves a 
non-degenerate quadratic form $\langle \cdot,\cdot\rangle$ on
$V$. First of all, this quadratic form induces an isomorphism 
$v\mapsto \langle v,\cdot \rangle$ between $V$ and $V^*$ which
intertwines the natural  
representation and its contragredient. So, there is no need in this
case to consider $V^*$. Then, if $\rho$ denotes as before the diagonal
action of $G$ on $V^{\otimes d}$, $\rho(\C G)'$ is larger
than\footnote{We keep the notation $\pi$ for the action of the
symmetric group of any order on the corresponding tensor power of $V$.}
$\pi(\S_d)$. The first fundamental theorem tells us how much larger.

In this section, we will identify freely $\End(V^{\otimes d})$ with
$V^{\otimes 2d}$ by saying that $v_1\otimes \ldots \otimes v_{2d}$
transforms $w_1\otimes \ldots \otimes w_d$ into $\prod_{i=1}^d \langle
v_i,w_i \rangle \; v_{d+1}\otimes \ldots \otimes v_{2d}$. 

Let $\tau$ be a partition of the set $\{1,\ldots,2d\}$ in pairs. Let
$(e_1,\ldots,e_m)$ be an orthonormal or standard basis of $V$,
according to the nature of $G$. We define $J_\tau
\in \End(V^{\otimes d})$ by
$$J_\tau=\sum_{i_1,\ldots,i_{2d}=1}^m \prod_{ \{k,l\} \in \tau, k< l} \langle
e_{i_k}, e_{i_l} \rangle \; e_{i_1}\otimes \ldots \otimes e_{i_{2d}}.$$
One checks easily that this definition of $J_\tau$ does not depend on
the choice of the orthonormal basis of $V$ and that $J_\tau$ commutes
to the action of $G$, that is, $J_\tau \in \rho(\C
G)'=\End_G(V^{\otimes d})$. 

\begin{figure}[h]
\begin{center}
\scalebox{0.7}{\includegraphics{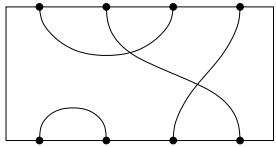}}
\caption{\label{j}The endomorphism $J_\tau$, when $d=4$ and
$\tau=\{\{1,3\},\{2,8\},\{4,7\},\{5,6\}\}$.} 
\end{center}
\end{figure}

The graphical representation introduced in the preceding section
may be helpful to clarify the situation. An example is given by Figure
\ref{j}. Note that we do not need arrows to distinguish between $V$ and $V^*$
anymore, since we are working in tensor powers of $V$ alone. 
 
The following theorem is proved in \cite{FH}.

\begin{theorem}[FFT for orthogonal and
symplectic groups] The
subspace \newline $\End_G(V^{\otimes d})=\rho(\C G)'$ of $\End(V^{\otimes d})$
is spanned by the endomorphisms $J_\tau$, where $\tau$ runs over the
partitions of $\{1,\ldots,2d\}$ in pairs.
\end{theorem}

\begin{remark} {\rm The proof of this theorem is longer than that of
Schur-Weyl duality, so we do not give it here. However, it is usually
stated and proved for complex Lie groups rather than compact ones.
Let us explain how the former can be deduced from the latter. If $G$
is $O(n)$ (resp. $Sp(n)$), let us denote by $G_\C$ the group $O_n\C$
(resp. $Sp_{2n}\C$). 
Since $G$ is contained in $G_\C$, one needs just prove that
any $u\in \End_G(V^{\otimes d})$ is invariant by the whole
$G_\C$. Via 
the isomorphism $\End(V^{\otimes d})\simeq V^{\otimes 2d}\simeq
(V^*)^{\otimes 2d}$, we can think of $u$ as a polynomial, that we
denote by $\tilde u$, in $2d$ variables on $V$, homogeneous of degree
one in each variable, invariant under the action of $G$. This
means that, for every $v\in V^{\oplus 2d}$, the function $\tilde
u(\cdot\;  v): G_\C \lra \C$ which sends $g$ to $\tilde u(gv)$ is
constant on $G$. Since, on one hand, this function is polynomial
in $g$ and on the other hand, $G$ is Zariski-dense in $G_\C$,
the function is constant on $G_\C$. So, $u$ is invariant by the
whole complex orthogonal group.

The theorem for $SO(n)$ follows from that for $O(n)$ just because
$\rho(\C SO(n))=\rho(\C O(n))$. } 
\end{remark}

We proceed now as before. Let $p_1,\ldots,p_r$ be integers. For each
$i=1\ldots r$, let $\alpha_i$ denote $V^{\otimes p_i}$ and set
$\balpha=(\alpha_1,\ldots,\alpha_r)$. Set $p=p_1+\ldots + p_r$. Let
$\tau$ be a partition of $\{1,\ldots,2r\}$ in pairs. 

\begin{proposition} \label{core-orthogonal} The spin network
$\psi_{\balpha,J_\tau}$ on $\L_r$ is a product of Wilson loops.
\end{proposition}

The proof is very similar to that of Proposition
\ref{core-unitary}. We are going to show that, up to some
isomorphism, $J_\tau$ acts as a permutation operator. For this, define
for each $i=1\ldots p$ $T_i=\pi((i,p+i)) \in\End(V^{\otimes 2p})
\simeq \End(\End(V^{\otimes p}))$. 

\begin{lemma} \label{j-sigma} Let $\tau$ be a partition of
$\{1,\ldots,2p\}$ in 
pairs. There exist $i_1,\ldots,i_k \in\{1,\ldots,p\}$ and $\sigma \in
\S_p$ such that $T_{i_1}\circ \ldots \circ
T_{i_k}(J_\tau)=\pi(\sigma)$.
\end{lemma}

\begin{remark}{\rm It is worth saying what this lemma means
graphically, because this is much simpler than the aspect of the proof
might suggest. Let us represent, as we did in Figure \ref{j}, a
partition like $\tau$ as a pairing of $2p$ points by $p$ lines. We
put the points $1,\ldots,p$ on the top edge of a box and
$p+1,\ldots,2p$ on the bottom edge, with $p+i$ below $i$. Then the
lemma says that, by switching the positions of $i$ and $p+i$ for some
well-chosen $i$'s without changing the pairing $\tau$, we can make sure that
every line connects a point on the top edge with a point on the bottom
edge. The diagram one gets in that way corresponds to a permutation
operator.}
\end{remark} 

\pf. It is convenient in this proof to think of $\tau$ as a
fixed-point free involution of $\{1,\ldots,2p\}$. Let
$\theta_1,\ldots,\theta_p$ denote the transpositions
$(1,p+1),\ldots,(p,2p)$. Then, given some integers $i_1,\ldots,i_k$
between $1$ and $p$, one checks easily that 
$$T_{i_1}\circ \ldots \circ T_{i_k}(J_\tau)=J_{\theta_{i_1}\ldots
\theta_{i_k} \tau \theta_{i_1}\ldots \theta_{i_k}},$$
where the product in the subscript of $J$ is a composition of
permutations of $\{1,\ldots,2p\}$. 
On the other hand, if $\sigma$ is a permutation of $\{1,\ldots,p\}$
and $\tau$ pairs $i$ with $\sigma(i)+p$ for each $i=1\ldots p$, then
$J_\tau=\pi(\sigma)$. 
Thus, the lemma will be proved if we show that, for some
$i_1,\ldots,i_k$ between $1$ and $p$ and some $\sigma\in\S_p$,
$\theta_{i_1}\ldots \theta_{i_k} \tau \theta_{i_1}\ldots
\theta_{i_k}$ pairs $i$ with $\sigma(i)+p$ for $i=1\ldots p$.

To do this, set $\theta=\theta_1\ldots \theta_p$. The map $\theta \tau$
acts on $\{1,\ldots,2p\}$ and we are interested in its orbits. In
particular, observe that $\{1,\ldots,p\}$ is a reunion of orbits of
$\theta \tau$ if and only if there exists $\sigma \in \S_p$ such that
$\tau$ pairs $i$ with $\sigma(i)+p$ for all $i=1\ldots p$. 

We define by induction a sequence of cycles, that is, of cyclic
permutations, on $\{1,\ldots,p\}$ as follows.

Set $x_1=1$ and let $\o_1$ be the orbit of $x_1$ under $\theta \tau$,
endowed with its cyclic order. Let $m_p:\{1,\ldots,2p\} \lra
\{1,\ldots,p\}$ be the map which sends $i$ and $i+p$ to $i$,
$i=1\ldots p$. Define the cycle $\c_1=m_p(\o_1)$ and set
$x_2=\min(\{1,\ldots,p\} - \c_1)$. Then, given $\c_1,\ldots,\c_{n-1}$
and $x_n$, define $\o_n$ as the orbit of $x_n$, $\c_n=m_p(\o_n)$ and
$x_{n+1}=\min(\{1,\ldots,p\} - (\c_1 \cup \ldots \cup \c_n))$. This
procedure stops when the cycles $c_1,\ldots,\c_n$ cover 
$\{1,\ldots,p\}$. These cycles are disjoint by construction and we see
each of them as an element of $\S_p$. Let us call $\sigma$ their product.

Now for each $i=1\ldots p$, one and only one of the two elements
$i$ and $i+p$ belong to $\o_1\cup \ldots \cup \o_n$. Set
$\epsilon_i=0$ if it is $i$, $\epsilon_i=1$ if it is $i+p$. Define
$\tilde \tau= (\prod_{i=1}^p \theta_i^{\epsilon_i}) \tau  (\prod_{i=1}^p
\theta_i^{\epsilon_i})$. It is easily checked that the iterates of
$\theta\tilde\tau$ preserve $\{1,\ldots,p\}$ and in fact that $\tilde\tau$
pairs $i$ with $\sigma(i)+p$ for $i=1\ldots p$. The lemma is
proved, by taking for $i_1,\ldots,i_k$ those integers $i$ such that
$\epsilon_i=1$. \qed

The elements of $G$ have a simple behaviour under the transposition
operators $T_i$. 

\begin{lemma} \label{transpo} Consider the following isomorphism:
$$T:\End(V)\simeq V^*\otimes V \build{-\!\!\!\lra}_{}^{\langle,\rangle
\otimes \id} V\otimes V \build{-\!\!\!\lra}_{}^{\pi((12))} V\otimes V
\build{-\!\!\!\lra}_{}^{\langle,\rangle \otimes \id} V^*\otimes
V\simeq \End(V).$$ 
Let $g$ be an element of $G$. Then $T(g)=\epsilon g^{-1}$, where
$\epsilon=1$ if $G$ is orthogonal and $\epsilon=-1$ if $G$ is
symplectic. 
\end{lemma} 

\pf. If $v$ belongs to $V$ and $\phi$ to $V^*$, let us denote by
$\tilde v$ and $\tilde \phi$ the corresponding elements of $V^*$ and
$V$ respectively, so that $\tilde v=\langle v,\cdot\rangle$ and
$\phi=\langle\tilde\phi , \cdot\rangle$. 

Let $g=\sum_i \phi_i \otimes v_i$ be an element of $G \subset
\End(V)$. Then $T(g)=\sum_i \tilde v_i \otimes \tilde\phi_i$. Now let
$u$ and $w$ be two elements of $V$. One has 
$$\langle gu,w\rangle = \langle \sum_i \phi_i(u) v_i,w\rangle 
= \sum_i \tilde v_i(w) \langle \tilde \phi_i,u\rangle 
= \epsilon \langle u, \sum_i \tilde v_i(w) \tilde\phi \rangle 
= \langle u, \epsilon T(g)(w)\rangle,$$
where $\epsilon$ equals plus or minus one, according to the symmetry
of the form $\langle\cdot,\cdot\rangle$. Since this form is
non-degenerate and preserved by $g$, the result follows.\qed 

\pf \textsc{ of Proposition \ref{core-orthogonal}. } Let
$g_1,\ldots,g_r$ be $r$ elements of $G$. By definition, 
$$\psi_{\balpha,J_\tau}(g_1,\ldots,g_r)=\tr\left(g_1^{\otimes p_1} \otimes
\ldots \otimes g_r^{\otimes p_r} \circ J_\tau\right).$$
By Lemma \ref{j-sigma}, there exist $i_1,\ldots,i_k \in
\{1,\ldots,p\}$ and $\sigma\in\S_p$ such that $T_{i_1}\circ \ldots \circ
T_{i_k}(J_\tau)=\pi(\sigma)$, or equivalently, $J_\tau=T_{i_1}\circ
\ldots \circ T_{i_k}(\pi(\sigma))$, since $T_i^2=1$. 

Now, observe that, for $u$ and $u'$ in $\End(V^{\otimes p})$, one has
$\tr(u \circ T_i(u'))=\tr(T_i(u) \circ u')$ for all $i=1\ldots
p$. Hence, we have
$$\psi_{\balpha,J_\tau}(g_1,\ldots,g_r)=\tr\left(T_{i_1}\circ \ldots \circ
T_{i_k}(g_1^{\otimes p_1} \otimes \ldots \otimes g_r^{\otimes p_r})
\circ \pi(\sigma)\right).$$
For the sake of clarity, let us rename the sequence
$(g_1,\ldots,g_1,\ldots,g_r,\ldots,g_r)$, where $g_i$ appears $p_i$
times, as just $(h_1,\ldots,h_p)$. Thus, $g_1^{\otimes p_1} \otimes
\ldots \otimes g_r^{\otimes p_r}$ equals $h_1\otimes \ldots \otimes
h_p$. Now, by Lemma \ref{transpo}, we have
$$T_{i_1}\circ \ldots \circ T_{i_k}(h_1\otimes \ldots \otimes h_p) =
\epsilon^k h_1^{\epsilon_1} \otimes \ldots \otimes h_p^{\epsilon_p},$$
where now $\epsilon_i=-1$ if $i$ appears in the list $i_1,\ldots,i_k$
and $\epsilon_i=1$ otherwise\footnote{This new $\epsilon_i$ is minus twice
the one defined in the proof of Lemma \ref{j-sigma}, plus one.}. Here again,
$\epsilon=1$ in the orthogonal case, $-1$ in the symplectic one. 

Now we finish the proof just as that of Proposition
\ref{core-unitary}. Indeed, 
\begin{eqnarray*}
\psi_{\balpha,J_\tau}(g_1,\ldots,g_r) &=& \epsilon^k \tr\left(h_1^{\epsilon_1}
\otimes \ldots \otimes h_p^{\epsilon_p} \circ \pi(\sigma) \right) \\
&=& \epsilon^k \prod_{\c=(a_1\ldots a_k)} \tr(h_{a_1}^{\epsilon_{a_1}} \ldots
h_{a_k}^{\epsilon_{a_k}}), 
\end{eqnarray*}
where the product runs over the decomposition of $\sigma$ in
cycles. Each factor in this product is a Wilson loop. Indeed, let us
define $j:\{1,\ldots,p\} \lra \{1,\ldots r\}$ by 
$$j(a)=\min \{i: a \leq p_1+\ldots +p_i \}.$$
Then by definition, $h_a=g_{j(a)}$. If $E=\{e_1,\ldots,e_r\}$ denotes
the set of edges of the graph $\L_r$, then we can define for every
cycle $\c=(a_1 \ldots a_k)$ of $\sigma$ the loop
$l_\c=(e_{j(a_1)}^{\epsilon_{a_1}},
\ldots,e_{j(a_k)}^{\epsilon_{a_k}})$. With this notation, we have proved that
$$ \psi_{\balpha,J_\tau}(g_1,\ldots,g_r) = \epsilon^k
\prod_{\c=(a_1\ldots a_k)} 
W_{n,l_\c},$$
where $n$ denotes the natural representation. This proves the
proposition. \qed

\bibliographystyle{plain}
\bibliography{wilson}

\end{document}